\documentclass[%
 reprint,
 superscriptaddress,
 bibnotes,
 amsmath,amssymb,
 aps,
]{revtex4-1}

\usepackage{graphicx}
\usepackage{subfigure}
\usepackage{dcolumn}
\usepackage{bm}
\usepackage[colorlinks, linkcolor=blue, urlcolor=blue, citecolor=blue]{hyperref}
\usepackage{float}
\usepackage[utf8]{inputenc}
\usepackage{epstopdf}
\begin{document}

\preprint{APS/123-QED}

\title{Anisotropic Andreev reflection in semi-Dirac materials}

\author{Hai Li}
\email{hnnuhl@hunnu.edu.cn}
\affiliation{Key Laboratory of Low-Dimensional Quantum Structures and Quantum Control of Ministry of Education, Key Laboratory for Matter Microstructure and Function of Hunan Province, School of Physics and Electronics, Hunan Normal University, Changsha 410081, China}

\author{Xiang Hu}
\affiliation{Key Laboratory of Low-Dimensional Quantum Structures and Quantum Control of Ministry of Education, Key Laboratory for Matter Microstructure and Function of Hunan Province, School of Physics and Electronics, Hunan Normal University, Changsha 410081, China}
\affiliation{Department of Physics, William $\&$ Mary, Williamsburg, VA 23187, USA}

\author{Gang Ouyang}
\email{gangouy@hunnu.edu.cn}
\affiliation{Key Laboratory of Low-Dimensional Quantum Structures and Quantum Control of Ministry of Education, Key Laboratory for Matter Microstructure and Function of Hunan Province, School of Physics and Electronics, Hunan Normal University, Changsha 410081, China}

\date{\today}

\begin{abstract}
 In the framework of Bogoliubov-de Gennes equation, we theoretically study the Andreev reflection in normal-superconducting junctions based on semi-Dirac materials. Owing to the intrinsic anisotropy of semi-Dirac materials, the configuration of Andreev reflection and differential conductance are strongly orientation-dependent. For the transport along the linear dispersion direction, the differential conductance exhibits a clear crossover from retro Andreev reflection to specular Andreev reflection with an increasing bias-voltage, and the differential conductance oscillates without a decaying profile when the interfacial barrier strength increases. However, for the transport along the quadratic dispersion direction, the boundary between the retro Andreev reflection and specular Andreev reflection is ambiguous, and the differential conductance decays with increasing the momentum mismatch or the interfacial barrier strength.  We illustrate the pseudo-spin textures to reveal the underling physics behind the anisotropic coherent transport properties. These results enrich the understanding of the superconducting coherent transport in semi-Dirac materials.

\end{abstract}

\maketitle

\section{\label{sec:level1} Introduction}

  Semi-Dirac material (SDM) has recently been experimentally realized in black phosphorus with in situ deposition of K \cite{Kim2015} or Rb \cite{Kim2017} atoms, offering a promising playground for further exploring the exotic attributes of SDMs. Unlike most Dirac materials that possess liner dispersions in all momentum-space directions \cite{Neto2009, Wehling, Armitage}, in SDMs the low-energy excitations disperse quadratically in one direction but linearly along the orthogonal direction \cite{ Pardo, Zhong, Pardo, Huang2015, Dietl, Delplace}. The unique band structures of SDMs are responsible for a series of novel phenomena \cite{Dietl, Delplace, Saha, Banerjee, Nualpijit, Mawrie, Adroguer, Islam, Mawrie2, Carbotte, Real, Chen, Zhai, Wang, Saha2016, Dutreix}, including the consequences of anisotropic aspect in the superconducting order parameter correlations \cite{Wang2019, Uchoa, Uryszek}. Recent theoretical efforts have demonstrated that the superconductivity in SDMs can be induced by arbitrarily weak attractions in the present of random chemical potential \cite{Wang2019}. Resorting to the mean-field calculation \cite{Uchoa} and renormalization group analysis \cite{Uryszek}, it is revealed that the s-wave superconductivity is more favorable in SDMs. More strikingly, owing to the intrinsic anisotropy, the stiffness of superconducting order parameter and the divergence behavior of correlation length are highly orientation-dependent \cite{Uchoa, Uryszek}. These progresses, together with the developments in the materialization of SDMs, provide foundations for exploring coherent transport properties in SDM-based normal-superconducting (NS) junctions.

  \par
  In a NS junction with ideal contacts, the transport properties are dominated by the Andreev reflection (AR) in the subgap energy regime of $E \le \Delta_0$, with $E$ the incident energy and $\Delta_0$ the superconducting gap \cite{Andreev, Pannetier}. In most conventional-metal-based NS junctions, the chemical potential in the N region satisfies $\mu_N \gg \Delta_0$, and the AR is a intra-band phase-coherent scattering process, during which an incident electron from the N region is reto-reflected as a hole \cite{Andreev, Pannetier, Zhu, Blonder}. While in the NS junctions based on Dirac materials, $\mu_N$ can be continuously tuned to the subgap regime satisfying $\mu_N < E$ \cite{Neto2009, Wehling, Beenakker2008}. Consequently, a conduction-band electron incident from the N region is specularly reflected back as a valence-band hole, leading to a inter-band phase-coherent scattering process known as specular-AR \cite{ Beenakker2008, Beenakker2006, Bhattacharjee, Linder, Li2016, Ludwig, Efetovprb}. Remarkably, by increasing $E$ within the subgap regime, a crossover from reto-AR to specular-AR occurs, manifesting itself as a dip at $E=\mu_N$ in the $E$-dependent conductance spectrum. This signature has been experimentally observed in bilayer-graphene-based NS junctions \cite{Efetov2016}. Moreover, in Dirac-material-based NS junction, due to the novel momentum-spin/pseudo-spin textures of Dirac fermions, the subgap differential conductance oscillates with the interfacial barrier strength without a decay profile \cite{Bhattacharjee}.

  \par
  Although the scenarios of AR in the systems with pure quadratic \cite{Andreev, Pannetier, Zhu, Blonder} or linear \cite{ Beenakker2006, Bhattacharjee, Linder, Li2016, Beenakker2008, Ludwig, Efetovprb} dispersions have triggered extensive studies, the AR and related subgap conductance in SDM-based NS junctions have received no attention to date. Since the low-energy excitations in SDMs host unique dispersions intermediate between the quadratic and linear energy spectra, it is natural to ask how the intrinsic anisotropy manifests itself in the superconducting coherent transport. In this paper we investigate the subgap transport properties in SDM-based NS junctions. The manifestations of the anisotropic dispersion in the subgap transport can be summarized as two points. First, we find a clear crossover from retro-AR to specular-AR when the transport along the linearly dispersion direction, while for the transport along the quadratical dispersion direction, the boundary between retro-AR and specular-AR is ambiguous. Second, the influences of momentum-mismatch and interfacial barrier on the subgap transport are strongly orientation-dependent. For the transport along the quadratic dispersion direction, the subgap differential conductance rapidly decays with increasing the interfacial barrier strength or momentum-mismatch between the N and S regions. Whereas the transport in the linear dispersion direction is insensitive to the momentum-mismatch, and the subgap differential conductance periodically oscillates with the interfacial barrier strength without a decaying profile. We illustrate the pseudo-spin textures of semi-Dirac fermions to understand our findings.

  \par
  This paper is organized as follows. We present the model and method in Sec.~\ref{sec:level2}. In Sec.~\ref{sec:level3}, we give the numerical results and concentrate on the manifestations of intrinsic anisotropy of SDMs in the subgap transport properties. The conclusions are briefly drawn in Sec.~\ref{sec:level4}.

\section{\label{sec:level2}  Model and Method}

  \begin{figure}
  \centering
  \includegraphics[width=8cm]{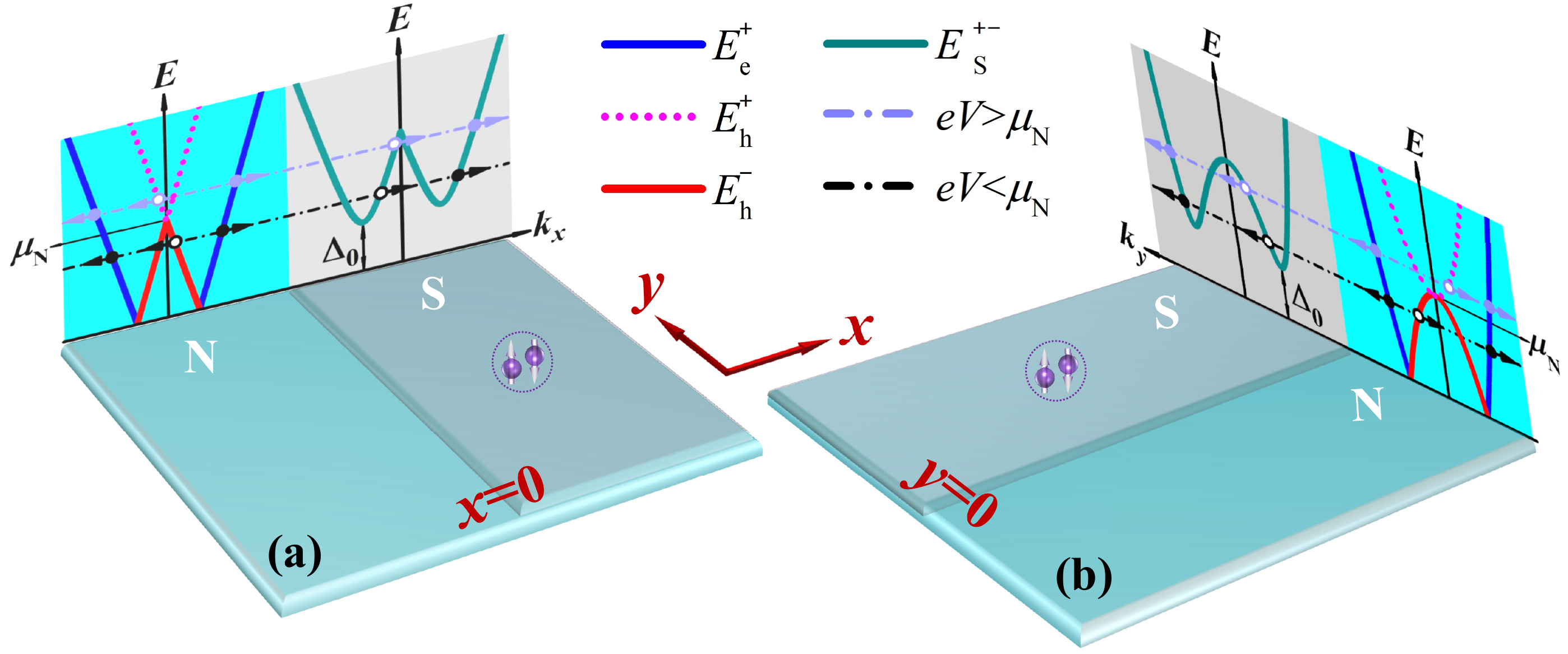}
  \caption{(Color online) Schematic plots of SDM-based NS junctions extending along (a) the $x$-direction and (b) the $y$-direction, and the low-energy excitations in the N regions disperse linearly and quadratically along the $x$- and $y$-directions, respectively. We assume the chemical potential in the N and S regions could be tuned independently. In the N region, $E^+_e$ denotes the electron-like conduction band, and $E^{+(-)}_h$ represents the hole-like valence (conduction) band. In the S region, $E^{+-}_S$ indicates one branch of Bogoliubov quasiparticle dispersion. The solid (open) circles denote electron (hole)-like quasiparticles and the arrows indicate the directions of group velocities.}
  \label{fig:fig1}
  \end{figure}

  \par
  To address the effects of the intrinsic anisotropy on the subgap transport properties, we introduce two representative SDM-based NS junctions extending along the $x$- and $y$- axes, respectively. As schematically shown in Fig.~\ref{fig:fig1}, the low-energy dispersion is linear (quadratic) in the N region of the NS junction extending along the $x(y)$-axis.  We consider the transport properties along the $x(y)$-direction in the NS junction extending along the $x(y)$-axis, and assume that the translational symmetry in the $y(x)$-direction is preserved, so that the transverse momentum $k_y(k_x)$ can be treated as a good quantum number \cite{Beenakker2006, Beenakker2008, Bhattacharjee, Ludwig, Linder, Li2016, Efetovprb}. Moreover, under this assumption the influences of boundary effects on the transport can be rationally neglected. In the S region, we take the intra-sublattice/orbit s-wave pairing, as proposed by recent theoretical work \cite{Uchoa, Uryszek}. In practice, the superconductivity in the S region can be induced by a s-wave superconductor via the proximity effect, as implemented in similar NS junctions based on graphene \cite{Heersche2007, Efetov2016} and Weyl semimetals \cite{Bachmanne, Huang2018}.

  \par
  Under these lines, the Bogoliubov–de Gennes (BdG) Hamiltonian describing the low-lying physics is given by \cite{Uchoa, Uryszek}
  \begin{equation}
  H_{\mathrm{BdG}} = \left( {\begin{array}{*{20}c}
  { h(\textbf{\emph{k}}})-\mu(\textbf{\emph{r}}) & { \Delta(\textbf{\emph{r}})} \\
  { \Delta^{\dag}(\textbf{\emph{r}})} & {- h(\textbf{\emph{k}}})+\mu(\textbf{\emph{r}}) \\
  \end{array}} \right),
  \label{eq:BdG}
  \end{equation}
  acting on the pseudo-spin $\otimes$ Nambu space. The single-particle effective Hamiltonian $ h(\textbf{\emph{k}})= \hbar v k_x  \sigma_x  + \eta  k^2_y \sigma_y$, where $\sigma_{x, y}$ label the Pauli matrices operating on the pseudo-spin space, $v$ represents the Fermi velocity, and $\eta \equiv \hbar^2/(2m)$ with $m$ the effective mass. The chemical potential $\mu(\textbf{\emph{r}})=\mu_N \Theta(-\textbf{\emph{r}}) + \mu_S \Theta(\textbf{\emph{r}})$, where $\Theta(\textbf{\emph{r}})$ is the Heaviside step function and $\mu_S$ ($\mu_N$) denotes the chemical potential in the S (N) region. In this paper, we assume that the relation of $\mu_S \gg \mu_N$ is satisfied, so that the leakage of Cooper pairs from the S to N regions can be safely neglected \cite{Beenakker2006, Bhattacharjee, Ludwig, Linder, Li2016, Efetovprb} . In doing so, the pair potential can be effectively modeled by a step function, i.e., $\Delta(\textbf{\emph{r}})=\Delta_0 \sigma_0 e^{i \phi} \Theta(\textbf{\emph{r}})$, with $\phi$ the superconducting phase and $\sigma_0$ a $2\times2$ identity matrix operating on the pseudo-spin space. The BdG Hamiltonian shown in Eq.~(\ref{eq:BdG}) can also be derived from the lattice model of a graphene-like system in proximity to a s-wave superconductor. The related calculation details are given in Appendix \ref{sec:levelappenA}.

  \par
  By diagonalizing the Hamiltonian shown in Eq.~(\ref{eq:BdG}) straightforwardly, the energy dispersion in the S region can be written as
  \begin{equation}
  E^{\nu \lambda}_S=\nu \sqrt{\left( \sqrt{\hbar^2 v^2 k_x^2+\eta^2 k^4_y } + \lambda \mu_S \right)^2+\Delta_0^2},
  \label{eq:ES}
  \end{equation}
  where $\nu=\pm$ and $\lambda=\pm$ indicate the different branches. To ensure the validity of mean-field approximation, the relation of $\Delta_0 \ll \mu_S$ should be satisfied. Therefore, the branches $E^{\nu +}_S$ are far away from the subgap regime and only $E^{\nu -}_S$ bands need to be considered. The schematic plots of $E^{+ -}_S$ as functions of $k_x$ and $k_y$ are shown in Fig.~\ref{fig:fig1} (a) and (b), respectively. In the N region, the low-energy spectrum can be formulated as
  \begin{equation}
  E^\pm_{e(h)}=\pm \sqrt{\hbar^2 v^2 k^2_x + \eta^2 k^4_y}-(+)\mu_N,
  \label{eq:EN}
  \end{equation}
  where the subscripts $e$ and $h$ denote the electronlike and holelike excitation spectra, respectively.

  \par
  According to Eq.~(\ref{eq:EN}), the group velocities in the N region can be parameterized as
  \begin{equation}
  v_{e(h), x} = \frac{\hbar v^2 k_x}{\varepsilon_{+(-)}},  v_{e(h), y} = \frac{2 \eta^2 k^3_y }{\hbar \varepsilon_{+(-)}},
  \label{eq:gv}
  \end{equation}
  where $\varepsilon_\pm = E \pm \mu_N$. The components $v_{e(h), x}$ and $v_{e(h), y}$ exhibit distinct behaviors with respect to the momentum, reflecting the anisotropic aspect of SDMs. We note that due to the intrinsic anisotropy, in the scattering issues the scattering angle should be defined as one between the directions of associated group velocity and current, instead of the angle between the directions of momentum and current. Resorting to Eq.~(\ref{eq:gv}), for a NS junction extending along the $y$-axis, the scattering angle should be $\alpha^y_{e(h)} \equiv \arctan (v_{e(h), x}/v_{e(h), y})=\hbar^2 v^2 k_x/(2 \eta^2 k^3_y)$, differing from the angle of $\arctan (k_x/k_y)$ in isotropic systems \cite{Beenakker2006, Bhattacharjee, Linder, Li2016}. For a NS junction extending along the $x$-axis, the scattering angle $\alpha^x_{e(h)}=\pi/2 -\alpha^y_{e(h)}$.

\subsection{\label{sec:level3-1} NS junction extending along the $x$-axis}

  We now turn to the scattering problem in a NS junction extending along the $x$-axis. In practice, at the boundary between the N and S regions, the defects and lattice mismatch are inevitable during the fabrication of device, which may profoundly influence the transport properties. To take into account the proposed effects, we introduce a interfacial barrier $U(x)\equiv U_0 \Theta(x)\Theta(d-x)$, and take the limit of $U_0 \rightarrow \infty$ and $d \rightarrow 0$ with $U_0d/(\hbar v)\equiv Z$ being finite. We note that in the current direction, the low-energy excitations in SDMs are similar as that in Dirac materials, thus the scattering amplitudes can be calculated by the standard procedure employed in related Dirac-material-based NS junctions \cite{Beenakker2006, Beenakker2008, Bhattacharjee, Ludwig, Linder, Li2016, Efetovprb}. By solving
  \begin{equation}
  t_{eq}  \psi^+_{eq} + t_{hq}  \psi^+_{hq} ={\cal M} ( \psi^+_e + r_{ee}  \psi^-_e + r_{he}  \psi^-_h )
  \label{eq:bcx}
  \end{equation}
  at the boundary of $x=0$, we can obtain $r_{he(ee)}$ and $t_{eq(hq)}$, which are the reflection amplitude for the AR (normal reflection) and the transmission amplitude for the electron(hole)-like quasiparticle, respectively. The details of basis scattering states $\psi^+_{eq(hq)}$ and $\psi^\pm_{e, h}$ are, respectively, given by Eqs.~(\ref{eq:wfsx}) and ~(\ref{eq:wfnx}). The transfer matrix ${\cal M}=e^{i \sigma_x \tau_0 Z}$, with $\tau_0$ a $2\times2$ identity matrix operating on the Nambu space.

  \par
  In the limit regime of $\mu_S \gg \max (\mu_N, E, \Delta_0)$, the reflection amplitudes are, respectively, given by
  \begin{subequations}
  \begin{equation}
  r_{ee}=\frac{ \hbar v (k^-_e \varepsilon_- \!-\! k^+_h \varepsilon_+)\cos \beta \!+\! i (\hbar^2 v^2 k^-_e k^+_h \!-\! \varepsilon_+ \varepsilon_-)\sin \beta}
  {\hbar v(k^+_h \varepsilon_+ \!+\! k^+_e \varepsilon_-)\cos \beta \!+\! i (\hbar^2 v^2 k^+_e k^+_h \!+\! \varepsilon_+ \varepsilon_-) \sin \beta},
  \end{equation}
  \begin{equation}
  r_{he}=\frac{2 \hbar v k_e \varepsilon_+ e^{-i \phi}}
  {\hbar v(k^+_h \varepsilon_+ \!+\! k^+_e \varepsilon_-)\cos \beta \!+\! i (\hbar^2 v^2 k^+_e k^+_h \!+\! \varepsilon_+ \varepsilon_-) \sin \beta},
  \end{equation}
  \label{eq:analyticr}
  \end{subequations}
  where $k^\pm_{e(h)} = k_{e(h)} \pm i \eta k^2_y/(\hbar v)$, $ k_{e(h)}={\mathrm{sgn}[\varepsilon_{+(-)}]}\sqrt{\varepsilon_{+(-)}^2-\eta^2 k^4_y}/(\hbar v)$, and $\beta = \cos^{-1}(E/\Delta_0)\Theta(\Delta_0-E)-i \cosh^{-1}(E/\Delta_0)\Theta(E-\Delta_0)$. As can be seen, both $r_{ee}$ and $r_{he}$ are independent of $Z$, implying that in the limit of $\mu_S \gg \max (\mu_N, E, \Delta_0)$, the transport properties along the $x$-axis direction are insensitive to the interfacial barrier.

  \par
  Resorting to the reflection amplitudes, the probabilities for the normal reflection (NR) and AR processes can be, respectively, defined as
  \begin{subequations}
  \begin{equation}
  R_{ee}=\left|\frac{\langle \psi^-_e | {\cal J}_x | \psi^-_e \rangle}
  {\langle \psi^+_e | {\cal J}_x | \psi^+_e \rangle}\right| |r_{ee}|^2,
  \end{equation}
  \begin{equation}
  R_{he}=\left|\frac{\langle \psi^-_h | {\cal J}_x | \psi^-_h \rangle}
  {\langle \psi^+_e | {\cal J}_x | \psi^+_e \rangle}\right| |r_{he}|^2,
  \end{equation}
  \label{eq:RX}
  \end{subequations}
  where the particle current density operator ${\cal J}_x \equiv \frac{-i}{\hbar}[x, H_{\mathrm{BdG}}]=v \sigma_x \tau_z$, with $\tau_z$ the Pauli matrix operating on the Nambu space.

  \par
  Taking advantage of the Blonder-Tinkham-Klapwijk (BTK) formula \cite{Blonder}, the zero-temperature differential conductance along the $x$-direction is given by
  \begin{equation}
  G^x(eV)=\frac{2e^2 W}{h} \! \int \! {\frac{d k_y}{2\pi} \left[1 \!-\! R_{ee}(eV, k_y) \!+\! R_{he}(eV, k_y)\right]},
  \label{eq:GX}
  \end{equation}
  where $V$ indicates the bias voltage and $W$ is the width along the $y$-direction of the junction. To normalize the conductance, we define $G^x_0(eV)=\frac{2e^2W}{\pi h}\sqrt{|eV+\mu_N|/\eta}$, which is the conductance along the $x$-direction of a SDM-based NN junction in the ballistic limit.

\subsection{\label{sec:level3-2} NS junction along the $y$-axis}

  In a NS junction along the $y$-axis, the low-energy excitations of the N region disperse quadratically in the current direction. For electron(hole)-like excitations with $|k_x|\le |E+(-)\mu_N|/(\hbar v)$,  Eq.~(\ref{eq:EN}) determines four possible scattering modes, including two propagating modes with $k_y = \pm q_{e(h)}$ and two evanescent ones with $k_y = \pm i \kappa_{e(h)}$. The related parameters are given by
  \begin{subequations}
  \begin{equation}
  q_{e(h)}=s_{e(h)}\sqrt{\sqrt{\varepsilon_{+(-)}^2-\hbar^2 v^2 k^2_x}/\eta},
  \label{eq:ky}
  \end{equation}
  \begin{equation}
  \kappa_{e(h)}= \sqrt{\sqrt{\varepsilon_{+(-)}^2-\hbar^2 v^2 k^2_x}/\eta},
  \label{eq:kappay}
  \end{equation}
  \end{subequations}
  with $s_{e(h)}={\mathrm{sgn}}[\varepsilon_{+(-)}]$. Although the evanescent modes do not contribute to the current, they need to be included in the wave functions to get correct boundary conditions. With this in mind, for a propagating electron-like incident mode, the wave function in the N region should be parameterized as
  \begin{equation}
  \Phi_N = \varphi^+_{e, 1} + \tilde r_{ee, 1}  \varphi^-_{e, 1} + \tilde r_{ee, 2}  \varphi^-_{e, 2} + \tilde r_{he, 1}  \varphi^-_{h, 1} + \tilde r_{he, 2}  \varphi^-_{h, 2},
  \label{eq:wfNY}
  \end{equation}
  with $\tilde r_{ee(he), 1}$ and $\tilde r_{ee(he), 2}$ the reflection amplitudes of propagating and evanescent modes in the NR(AR) processes, respectively. The detailed structures of related basis scattering states are given by Eq.~(\ref{eq:wfny}). In the S region, the wave function takes the form of
  \begin{equation}
  \Phi_S = \tilde t_{eq, 1}  \varphi^+_{eq, 1} + \tilde t_{eq, 2}  \varphi^+_{eq, 2} +\tilde t_{hq, 1}  \varphi^+_{hq, 1} +\tilde t_{hq, 2}  \varphi^+_{hq, 2} ,
  \label{eq:wfSY}
  \end{equation}
  where the associated basis scattering states are shown in Eq.~(\ref{eq:wfsy}). The coefficients $\tilde t_{eq(hq), 1}$ and $\tilde t_{eq(hq), 2}$, respectively, represent the transmission amplitudes for the propagating and evanescent electron(hole)-like quasiparticles.

  \par
  To model the effects resulting from the interfacial defects and lattice mismatch, we introduce a interfacial barrier modeled by $\tilde U(y)\equiv \tilde U_0 \delta(y)$, with $\delta(y)$ being the Delta function. In doing so, the boundary conditions can be formulated as
  \begin{subequations}
  \begin{equation}
  \Phi_S|_{y =0^+}=\Phi_N|_{y=0^-},
  \end{equation}
  \begin{equation}
  \partial_y \Phi_S|_{y =0^+} = {\tilde {\cal M}} \Phi_N|_{y=0^-},
  \end{equation}
  \label{eq:bcY}
  \end{subequations}
  where ${\tilde {\cal M}}=\partial_y-\frac{\tilde U_0}{\eta}\sigma_y \tau_0$. For brevity of notation, we introduce a parameter $\chi \equiv \frac{\tilde U_0}{\eta q^F_S}$ to characterize the interfacial barrier strength, with $q^F_S=\sqrt{\mu_S/\eta}$. The probabilities for the NR and AR processes are, respectively, defined as
  \begin{subequations}
  \begin{equation}
  \tilde R_{ee, 1 (2)}=\left|\frac{\langle \varphi^-_{e, 1 (2)} | {\cal J}_y | \varphi^-_{e, 1 (2)} \rangle}
  {\langle \varphi^+_{e,1} | {\cal J}_y | \varphi^+_{e,1} \rangle}\right| |\tilde r_{ee, 1 (2)}|^2,
  \end{equation}
  \begin{equation}
  \tilde R_{he, 1 (2)}=\left|\frac{\langle \varphi^-_{h, 1 (2)} | {\cal J}_y | \varphi^-_{h, 1 (2)} \rangle}
  {\langle \varphi^+_{e,1} | {\cal J}_y | \varphi^+_{e,1} \rangle}\right| |\tilde r_{he, 1 (2)}|^2,
  \end{equation}
  \end{subequations}
  with the particle current density operator ${\cal J}_y \equiv \frac{-i}{\hbar}[y, H_{\mathrm{BdG}}]=-\frac{2i\eta}{\hbar} \sigma_y \tau_z \partial y$. We note that the probabilities of evanescent modes $\tilde R_{ee, 2}$ and $\tilde R_{he, 2}$ are vanishing and do not contribute to the current. Therefore, in accordance with the BTK formula \cite{Blonder}, the zero-temperature differential conductance along the $y$-direction can be written as
  \begin{equation}
  G^y(eV)=\frac{2e^2 L}{h}\! \int \! {\frac{d k_x}{2\pi} \left[1 \!-\! \tilde R_{ee, 1}(eV, k_x) \!+\! \tilde R_{he, 1}(eV, k_x)\right]},
  \label{eq:GY}
  \end{equation}
  where $L$ is the width along the $x$-direction of the junction. It is convenient to normalize the conductance by $G^y_0(eV)=\frac{2e^2L}{\pi h}\frac{|eV+\mu_N|}{\hbar v}$, which represents the conductance in the $y$-direction of a SDM-based NN junction in the ballistic limit.

  \begin{figure}
  \centering
  \includegraphics[width=8cm]{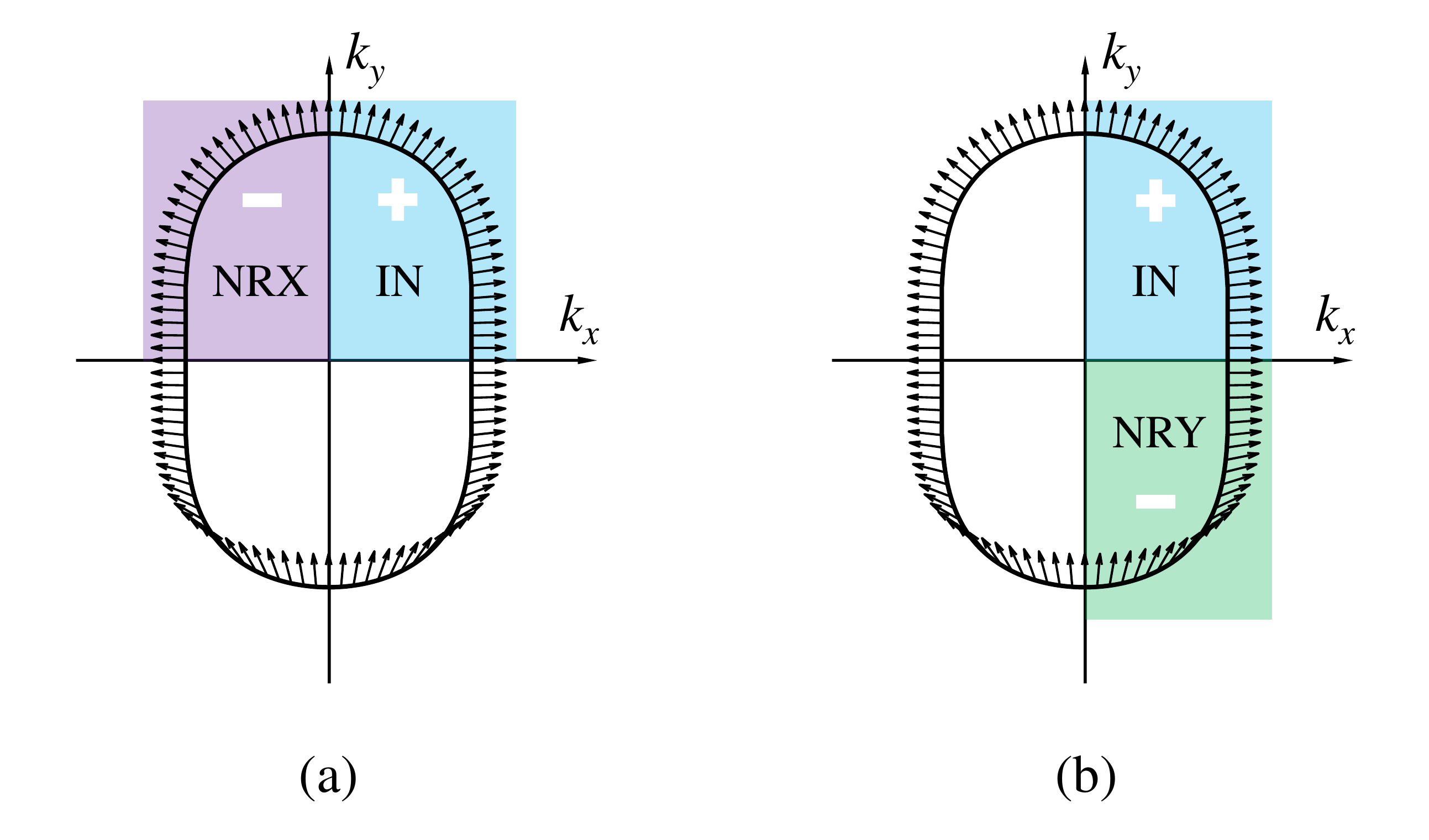}
  \caption{(Color online) Schematic plots of pseudo-spin texture of an electron-like conduction band in the N region of a SDM-based NS junction, where the arrow denotes the pseudo-spin and the solid curve depicts the iso-energy contour. The symbol $+$ ($-$) in (a) and (b), respectively, denotes the direction of $v_{e, x}$ and $v_{e, y}$ is parallel (anti-parallel) to the current direction. The notations IN and NRX (NRY) indicate the $\textbf{\emph{k}}$-space regions of incidence and normal reflection in the NS junction extending along the $x(y)$-axis.}
  \label{fig:fig2}
  \end{figure}

\section{\label{sec:level3} Results and Discussion}

\subsection{\label{sec:level3-1} Pseudo-spin Textures}

  \par
  To understand the underlying physics behind the transport properties, it is instructive to reveal the pseudo-spin textures in the N regions of SDMs-based NS junctions. To do so, we define the pseudo-spin vector as $\textbf{\emph{P}} \equiv \langle \bm{\sigma} \otimes \tau_0  \rangle$, which is a unit vector consisting of the expectation values of operator $\bm{\sigma} \otimes \tau_0$ in normalized basis scattering states, where $\bm \sigma \equiv (\sigma_x, \sigma_y)$ \cite{San-Jose, Schomerus, Majidi, Majidi, Habe}. Specifically, the pseudo-spin carried by a propagating electron-like mode can be formulated as
  \begin{equation}
  {\bm P}_e (k_x, k_y)= \left(  \frac{\hbar v k_x}{\varepsilon_{+}}, \frac{\eta k_y^2}{\varepsilon_{+}} \right).
  \label{eq:pseudospin}
  \end{equation}
  Accordingly, the pseudo-spin components $P_{e, x}$ and $P_{e, y}$ depend, respectively, linearly and quadratically on the momentum, inheriting the intrinsic anisotropy of SDMs. Fig.~\ref{fig:fig2} gives the schematic plots for the pseudo-spin textures of electron-like conduction band. As can be seen, $P_{e, y}$ is always along the positive $y$-direction, regardless of the sign of $k_y$, this scenario is profoundly different from that in Dirac materials \cite{San-Jose, Schomerus, Majidi, Majidi, Habe, Breunig, Uchida, Bai2020}. As will be proposed in Sec.~\ref{sec:level3-2}, the unique pseudo-spin textures provide illuminating elucidations for the subgap transport properties in SDMs-based NS junctions.

\subsection{\label{sec:level3-2} Differential conductance}

  \begin{figure*}
  \centering
  \subfigure{
  \includegraphics[width=7cm]{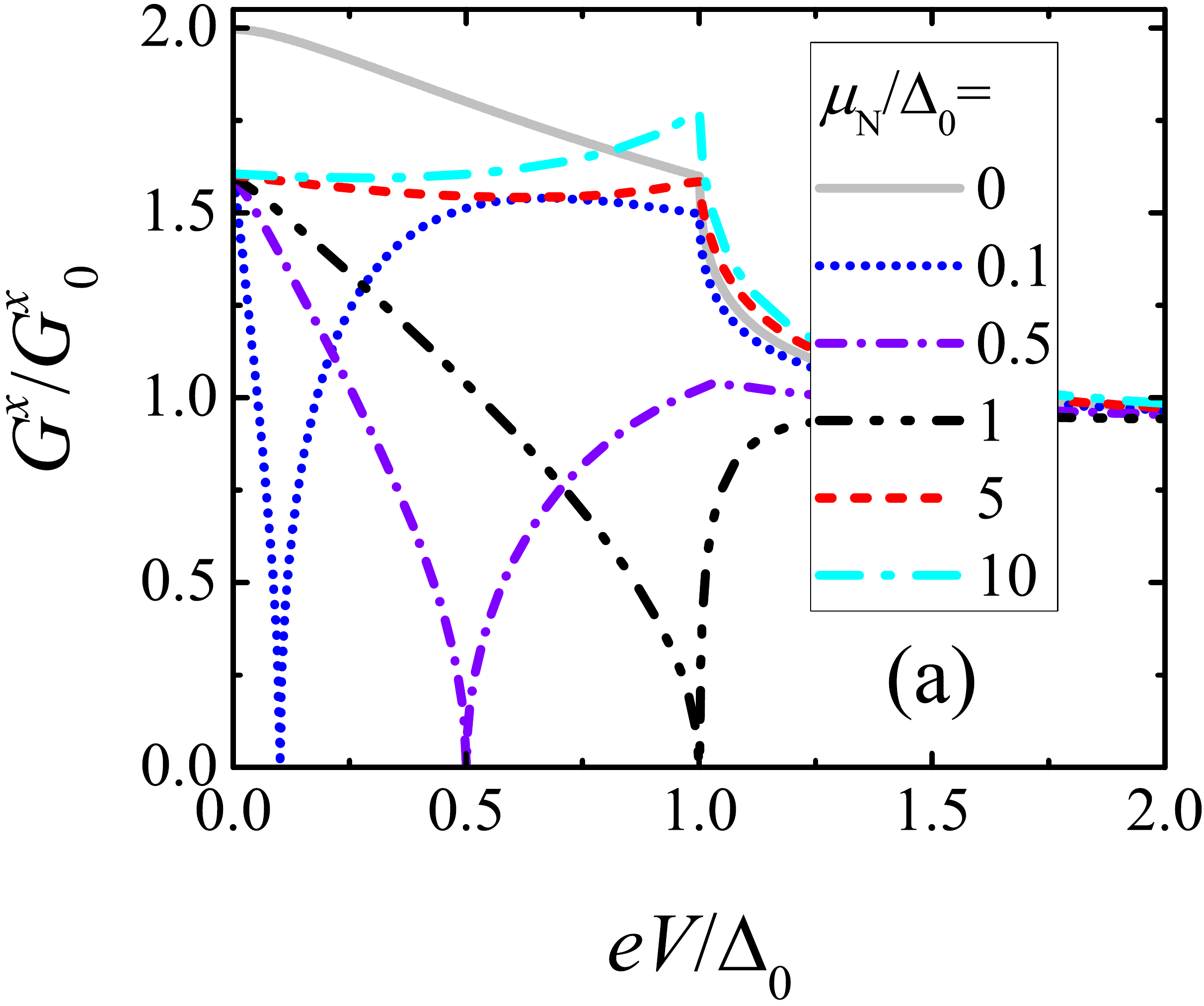}
  \label{subfig:fig3a}}
  \vspace{0.0cm}
  \hspace{2.0cm}
  \subfigure{
  \includegraphics[width=7cm]{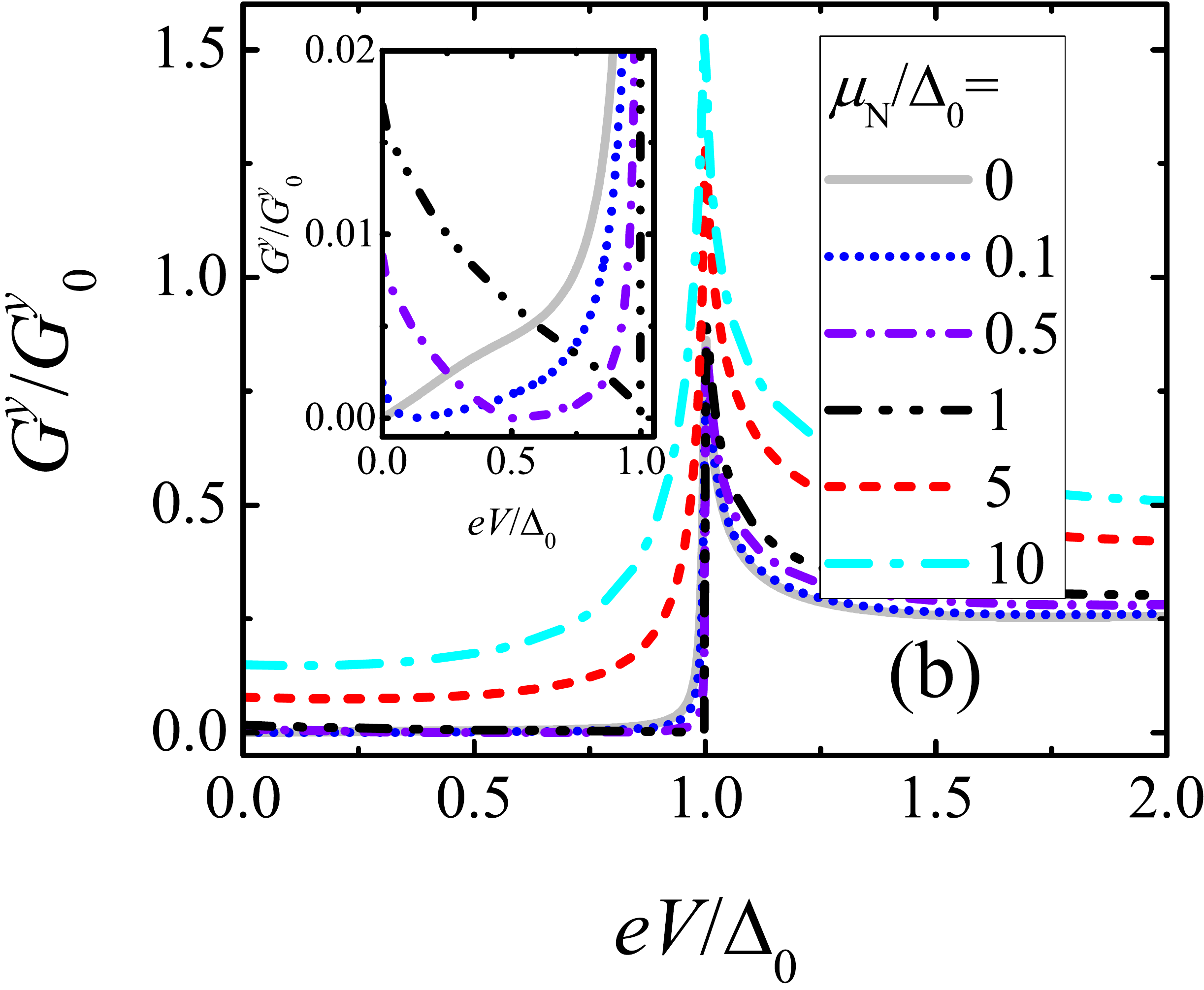}
  \label{subfig:fig3b}}
  \vspace{0.0cm}
  \hspace{0.0cm}
  \caption{(Color online) Bias-voltage-dependent differential conductance for NS junctions extending along the (a) $x$-direction and (b) $y$-direction, without interfacial barriers. The inset in panel (b) is the zoom-in of sub-gap conductance for $\mu_N/\Delta_0=0$, $0.1$, $0.5$, and $1$.}
  \label{fig:fig3}
  \end{figure*}

  In this subsection, we proceed to analyze the numerical results and discuss the differential conductance in the subgap regime. What we mainly concentrate on are the manifestations of the intrinsic anisotropy in the subgap transport properties, implemented by comparing the AR configurations and subgap differential conductance along the $x$- and $y$-axes. To ensure the validity of the mean-field approximation, we assume the S regions are heavily doped to the regime of $\mu_S \gg \Delta_0$, and we set $\mu_S=500\Delta_0$ in the numerical calculation for definiteness.

  \par
  As a starting point, we focus on the bias-voltage-dependent subgap differential conductance, and analyze the results in terms of the unique pseudo-spin textures. In a NS junction extending along the $x$-axis, the zero-bias conductance is insensitive to the momentum-mismatch (or alternatively, the ratio of $\mu_N/\mu_S$). As shown in Fig.~\ref{subfig:fig3a}, the differential conductance always exhibits a zero-bias peak, regardless of the value of $\mu_N$. Remarkably, for any nonvanishing $\mu_N$, $G^x(0)/G^x_0(0)$ takes the same value of $\frac{8}{5}$. On the contrary, the $G^y(0)/G^y_0(0)$ strongly depends on the momentum-mismatch in the NS junction extending along the $y$-axis. As can be seen in Fig.~\ref{subfig:fig3b}, the subgap differential conductance almost vanishes in the presence of large momentum-mismatch (e.g., $\mu_N/\Delta_0=0, 0.1, 0.5, 1$). Therefore, the behavior of zero-bias conductance is highly orientation-dependent. This scenario can be understood by virtue of the anisotropic pseudo-spin textures illustrated in Fig.~\ref{fig:fig2}. In the NS junction along the $x$-direction, for a small $k_y$ (i.e., with a small incident angle), the pseudo-spin carried by an electron-like incident mode is nearly opposite to that of the normally reflected one. Consequently, when $k_y$ is small enough, the conservation of pseudo-spin results in the prohibition of NR, and thus leads to the enhancement of AR in the subgap regime, even in the presence of strong momentum-mismatch. Moreover, since the conductance is mainly contributed by the modes with small $k_y$, the zero-bias conductance is insensitive to the momentum-mismatch. However, in the NS junction extending along the $y$-direction, a couple of incident and normally reflected modes possess the same pseudo-spin, so that the back scattering is enabled. Therefore, the subgap differential conductance in the $y$-direction is strongly suppressed by increasing the momentum-mismatch.

  \begin{figure*}
  \centering
  \subfigure{
  \includegraphics[width=7cm]{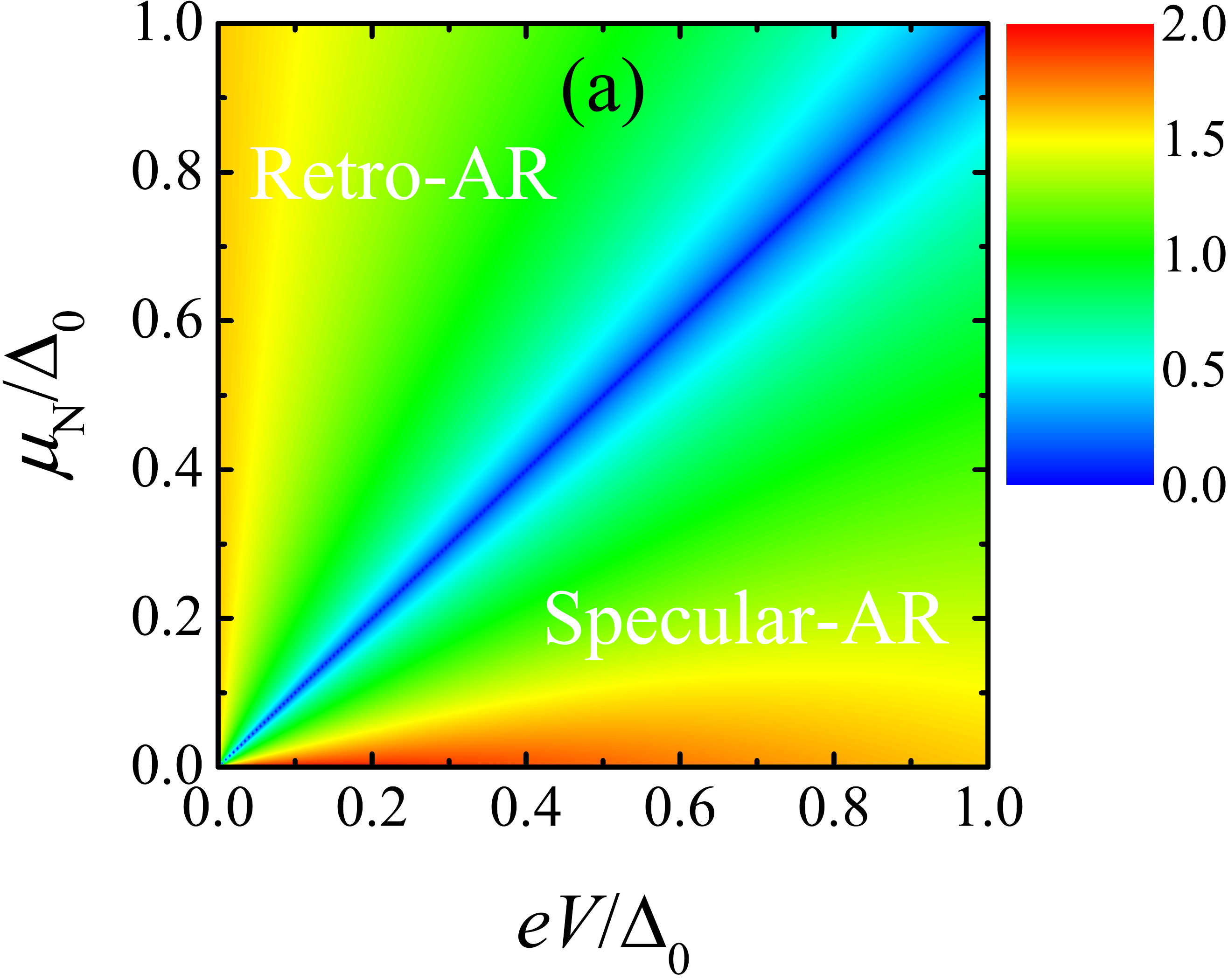}
  \label{subfig:fig4a}}
  \vspace{0.0cm}
  \hspace{2.0cm}
  \subfigure{
  \includegraphics[width=7.1cm]{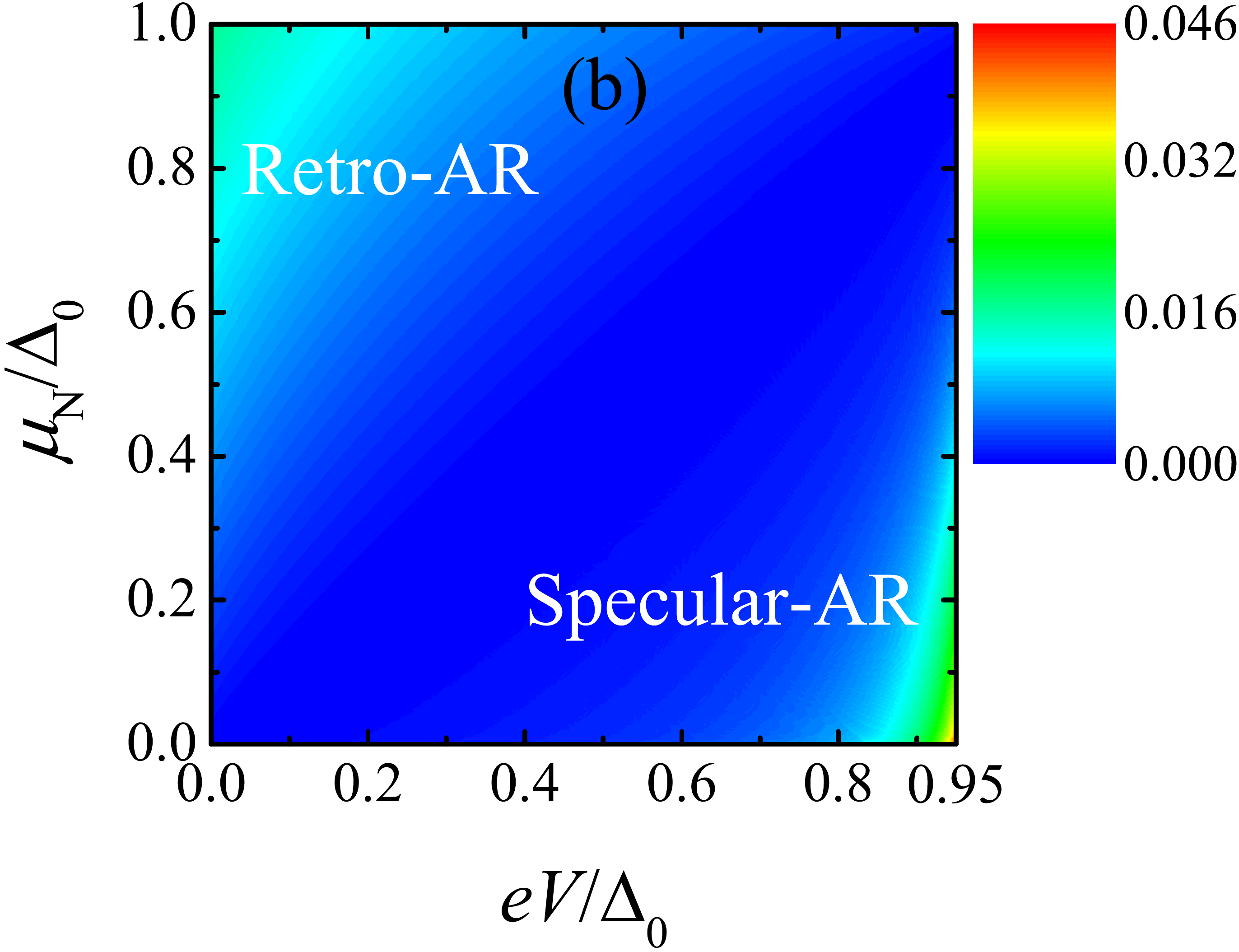}
  \label{subfig:fig4b}}
  \vspace{0.0cm}
  \hspace{0.0cm}
  \caption{(Color online) Panels (a) and (b), respectively, present the contour plots of differential conductance $G^x(eV)/G^x_0(eV)$ and $G^y(eV)/G^y_0(eV)$, without interfacial barriers. The differential conductance vanishes at the line of $eV=\mu_N$, indicating the crossover boundary between retro-AR and specular-AR.}
  \label{fig:fig4}
  \end{figure*}

  \par
  Parenthetically, although the configuration of $G^x(eV)/G^x_0(eV)$ shown in Fig.~\ref{subfig:fig3a} is similar as that in associated NS junctions based on graphene \cite{Beenakker2006} and silicene \cite{Linder, Li2016}, there is a significant difference on the value of $G^x(0)/G^x_0(0)$. For $eV=0$, according to Eqs.~(\ref{eq:analyticr}) and (\ref{eq:RX}), the reflection probabilities reduces into $R_{ee}|_{eV=0}=\eta^2 k^4_y / \mu^2_N$ and $R_{he}|_{eV=0}= (\mu^2_N - \eta^2 k^4_y)/\mu^2_N$, respectively. Substituting the results into Eq.~(\ref{eq:GX}), we arrive at $G^x(0)/G^x_0(0)=\sqrt{\frac{\eta}{\mu_N}}\int^{\sqrt{\mu_N/\eta}}_0 (2-2\eta^2k^4_y/\mu_N^2)dk_y =\frac{8}{5}$,  differing from the value of $\frac{4}{3}$ in graphene- and silicene-based NS junctions \cite{Beenakker2006, Linder, Li2016}. This consequence can be ascribed to the unique band structure of SDMs intermediate the linear and quadratic spectra.

  \begin{figure*}
  \centering
  \subfigure{
  \includegraphics[width=7cm]{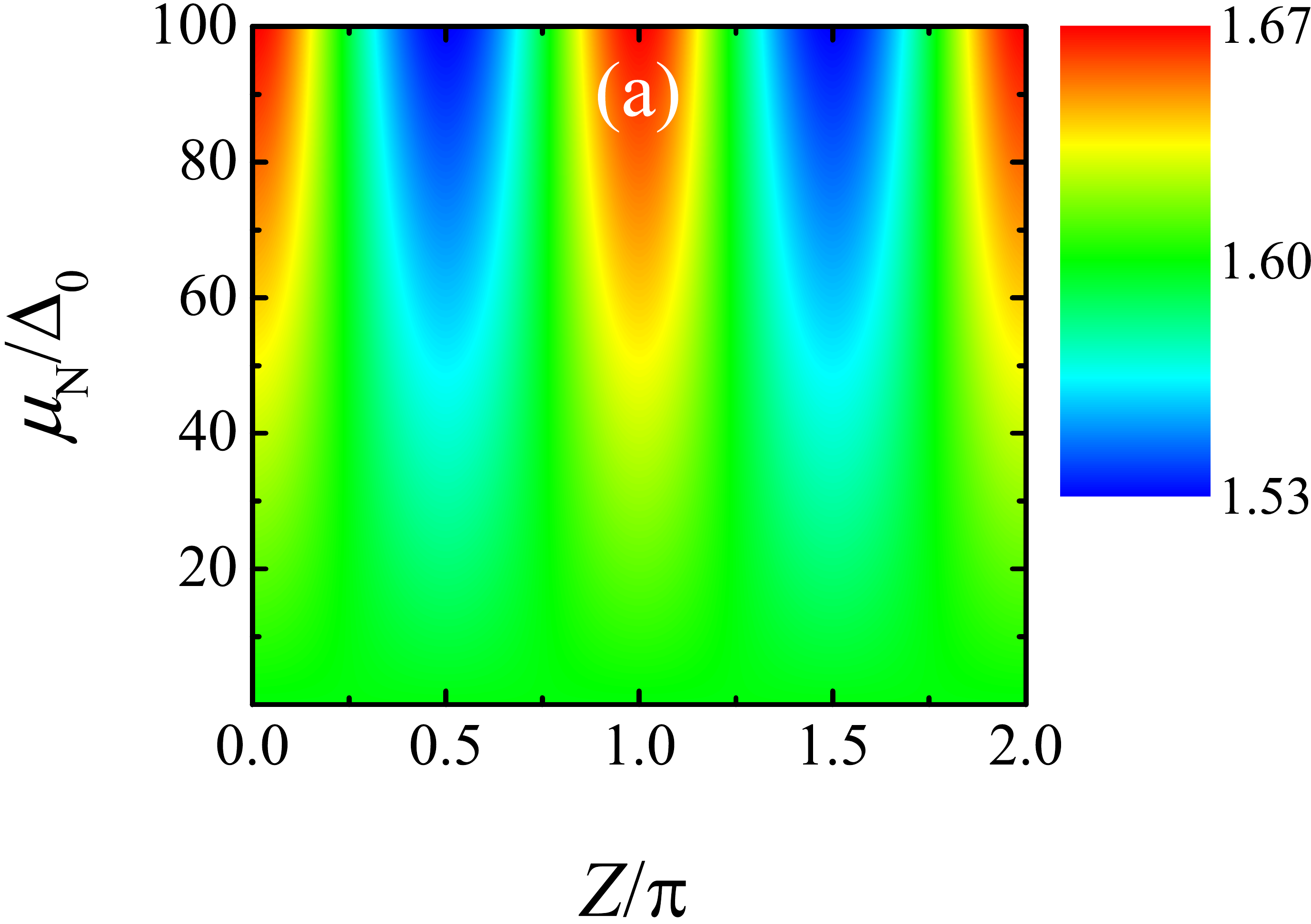}
  \label{subfig:fig5a}}
  \vspace{0.0cm}
  \hspace{2.0cm}
  \subfigure{
  \includegraphics[width=7cm]{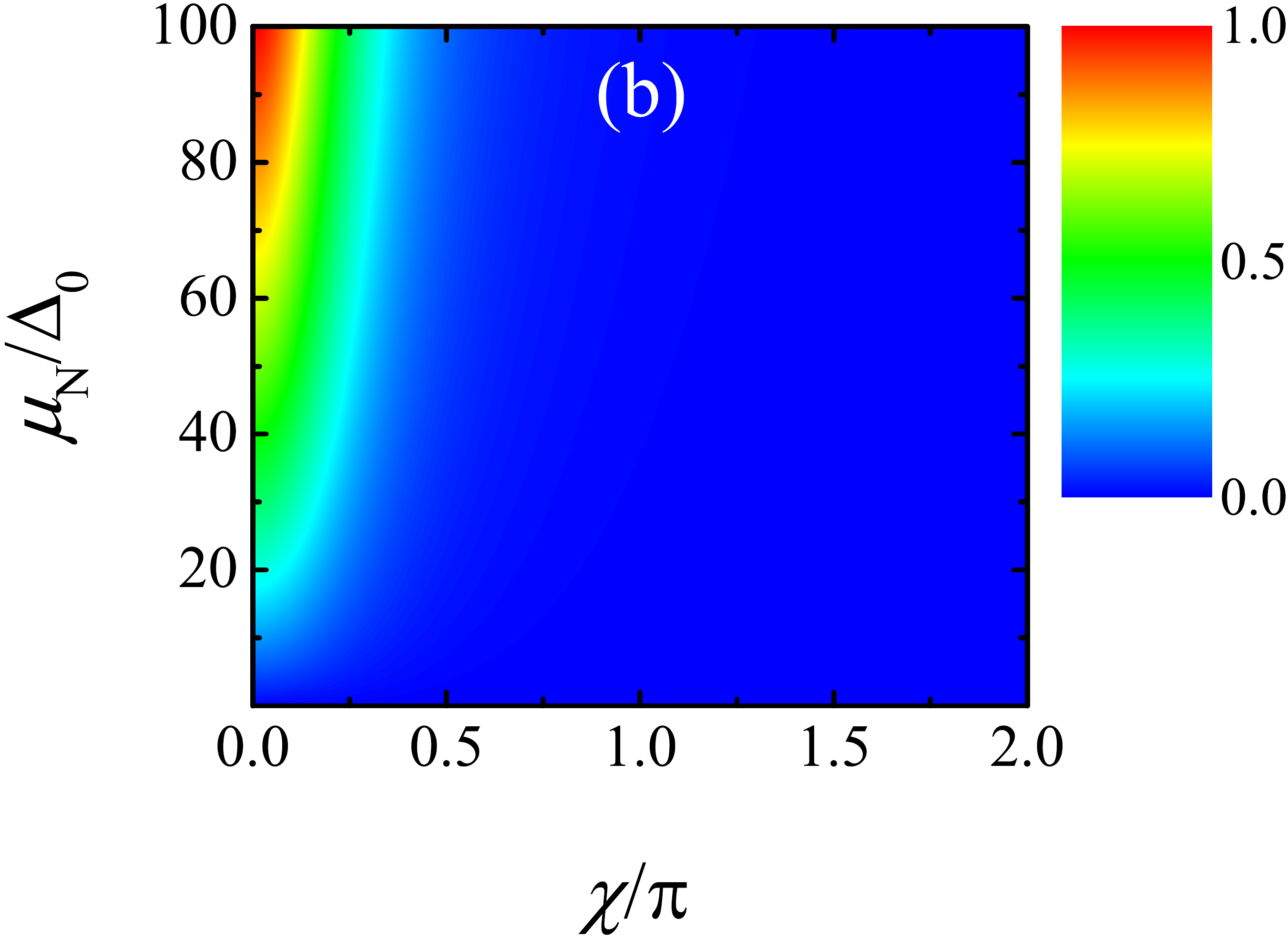}
  \label{subfig:fig5b}}
  \vspace{0.0cm}
  \hspace{0.0cm}
  \caption{ (Color online) Contour plots of the zero-bias differential conductance (a) $G^x(0)/G^x_0(0)$ and (b) $G^y(0)/G^y_0(0)$, where the starting value of $\mu_N$ is chosen as $10^{-3}\Delta_0$. }
  \label{fig:fig5}
  \end{figure*}

  \par
  The anisotropic aspect of the subgap differential conductance also manifests itself in the distinct crossover behaviors from the retro-AR to specular-AR in the two representative NS junctions. In the NS junction with a weakly doped N region satisfying $\mu_N < \Delta_0$, the subgap differential conductance vanishes at $eV=\mu_N$, due to the absence of AR for any incident angles. Furthermore, in the regimes of $eV<\mu_N$ and $eV>\mu_N$, the subgap differential conductance is dominated by the retro-AR and specular-AR, respectively. Therefore, the point of $eV=\mu_N$ serves as the crossover boundary between the retro-AR and specular-AR in the subgap regime. In the NS junction along the $x$-axis, as the bias-voltage increases, the subgap differential conductance exhibits a clear crossover from retro-AR to specular-AR, as depicted by Fig.~\ref{subfig:fig4a}. However, as illustrated in Fig.~\ref{subfig:fig4b}, the boundary between retro-AR and specular-AR is ambiguous in the NS junction extending along the $y$-axis. This scenario can be understood from the unique pseudo-spin textures. In the NS junction extending in the $y$-axis, the incident and normally reflected modes carry the same pseudo-spin, thus the NR is favorable, especially in the presence of large momentum-mismatch. Consequently, as illustrated in Fig.~\ref{subfig:fig3b} and the inset therein, for $\mu_N \le \Delta_0 \ll \mu_S$, the AR process is strongly suppressed so that $G^y(eV)/G^y_0(eV) \sim G^y(\mu_N)/G^y_0(\mu_N) =0$ in the subgap regime, thus blurring the boundary between the retro-AR and specular-AR.

  \par
  We now turn to the effects of the interfacial barriers on the subgap differential conductance. For the transport along the $x$-axis, the pseudo-spin carried by the incident and normally reflected modes are nearly opposite to each other when the incident angle is small enough, resulting in the prohibition of NR. As a consequence, the zero-bias differential conductance just oscillates with $Z$ without a decaying profile, as charted out by Fig.~\ref{subfig:fig5a}. In the case of the NS junction extending along the $y$-axis, the back scattering is available since the incident and normally reflected modes host the same pseudo-spin. Therefore, the AR is strongly suppressed by increasing the interfacial barrier strength, leading to the exponential decaying profile in the $\chi$-dependent $G^y(0)/G^y_0(0)$ shown in Fig.~\ref{subfig:fig5b}. In this regard, the influences of the interfacial barrier on the subgap transport are orientation-dependent, reflecting the anisotropic aspect of SDMs.

\section{\label{sec:level4} Conclusion}

  In conclusion, we have studied the subgap transport properties in SDMs-based NS junctions in the framework of BdG equation.  In terms of the tight-binding approach, the BdG Hamiltonian has been derived from a graphene-like system proximitized to a s-wave superconductor. We have figured out the manifestations of the intrinsic anisotropy of SDMs in the AR configurations and subgap differential conductance. For the transport along the linear dispersion direction, the subgap differential conductance exhibits a clear crossover from retro-AR to specular-AR by enhancing the bias-voltage, and the zero-bias differential conductance is insensitive to the momentum mismatch. Moreover, when the interfacial barrier strength increases, the zero-bias conductance exhibits an oscillating configuration without a decaying profile. However, for the transport along the quadratic dispersion direction, the boundary between the retro-AR and specular-AR is ambiguous, and the the zero-bias differential conductance rapidly drops as the momentum mismatch or the interfacial barrier strength increases. These results would provide some intriguing insights for the coherent transport in SDMs-based superconducting hybrid structures, and we anticipate more interesting results for the Andreev bound states and supercurrents in SDMs-based Josephson junctions.

\begin{acknowledgments}
 H. L. and X. H. would like to thank E. Rossi and  R. Wang for helpful discussions. This work was supported by the National Natural Science Foundation of China (Grants No 11804091 and No. 91833302), the Science and Technology Planning Project of Hunan Province (Grant No. 2019RS2033), the Hunan Provincial Natural Science Foundation of China (Grant No. 2019JJ50380), and the excellent youth fund of the Hunan Provincial Education Department (Grant No. 18B014).
\end{acknowledgments}


\appendix

\begin{widetext}

\section{\label{sec:levelappenA} Derivation of the BdG Hamiltonian}

  Physically, the semi-Dirac dispersion can be realized in a graphene-like systems with breaking the hexagonal symmetry \cite{Chen, Zhong, Mawrie2}. Thus we take an artificial honeycomb lattice model with anisotropic nearest-neighbor (NN) hopping as a prototype. The NN tight-binding model containing an intro-sublattice/orbit Bardeen-Cooper-Schrieffer superconducting pairing term is given by
  \begin{equation}
  H= \sum_{\emph{\textbf{l}}, \alpha} [(t B^\dag_{\emph{\textbf{l}}+\emph{\textbf{m}}_1, \alpha} A_{\emph{\textbf{l}}, \alpha}
                                       +t B^\dag_{\emph{\textbf{l}}+\emph{\textbf{m}}_2, \alpha} A_{\emph{\textbf{l}}, \alpha}
                                       +t^\prime B^\dag_{\emph{\textbf{l}}, \alpha} A_{\emph{\textbf{l}}, \alpha})+ h.c.]
                                       -\mu \sum_{\emph{\textbf{l}}, \alpha}(A^\dag_{\emph{\textbf{l}}, \alpha} A_{\emph{\textbf{l}}, \alpha}+B^\dag_{\emph{\textbf{l}}, \alpha} B_{\emph{\textbf{l}}, \alpha})
                                       +\sum_{\emph{\textbf{l}}, \alpha} [s \Delta (A^\dag_{\emph{\textbf{l}}, \alpha} A^\dag_{\emph{\textbf{l}}, -\alpha} + B^\dag_{\emph{\textbf{l}}, \alpha} B^\dag_{\emph{\textbf{l}}, -\alpha})+ h.c.]
  \end{equation}
  where the primitive lattice vectors $\emph{\textbf{m}}_1=\frac{a}{2}(3, \sqrt{3})$ and $\emph{\textbf{m}}_2=\frac{a}{2}(3, -\sqrt{3})$ with $a$ the interatomic distance, and $s=+(-)$ for $\alpha=\uparrow(\downarrow)$. Performing Fourier-transformations according to $a_{\emph{\textbf{k}}, \alpha}=\frac{1}{\sqrt N}\sum_{\emph{\textbf{l}}_A} A_{\emph{\textbf{k}}, \alpha} e^{-i \emph{\textbf{k}} \cdot \emph{\textbf{l}}_A}$ and $b_{\emph{\textbf{k}}, \alpha}=\frac{1}{\sqrt N}\sum_{\emph{\textbf{l}}_B} B_{\emph{\textbf{k}}, \alpha} e^{-i \emph{\textbf{k}} \cdot \emph{\textbf{l}}_B}$, we arrive at
  \begin{equation}
  H= t \sum_{\emph{\textbf{k}}, \alpha} [ \gamma_{\emph{\textbf{k}}} b^\dag_{\emph{\textbf{k}}, \alpha} a_{\emph{\textbf{k}}, \alpha} + \gamma^{\ast}_{\emph{\textbf{k}}} a^\dag_{\emph{\textbf{k}}, \alpha} b_{\emph{\textbf{k}}, \alpha}]
  - \mu \sum_{\emph{\textbf{k}}, \alpha} (a^\dag_{\emph{\textbf{k}}, \alpha} a_{\emph{\textbf{k}}, \alpha} + b^\dag_{\emph{\textbf{k}}, \alpha} b_{\emph{\textbf{k}}, \alpha})
  + \sum_{\emph{\textbf{k}}, \alpha} [s \Delta (a^\dag_{\emph{\textbf{k}}, \alpha} a^\dag_{-\emph{\textbf{k}}, -\alpha} + b^\dag_{\emph{\textbf{k}}, \alpha} b^\dag_{-\emph{\textbf{k}}, -\alpha}) + s \Delta^\ast (a_{-\emph{\textbf{k}}, -\alpha} a_{\emph{\textbf{k}}, \alpha} + b_{-\emph{\textbf{k}}, -\alpha} b_{\emph{\textbf{k}}, \alpha})],
  \label{eq:Hk}
  \end{equation}
  where the parameter $\gamma_{\emph{\textbf{k}}}$ is given by
  \begin{equation}
  \gamma_{\emph{\textbf{k}}} = e^{-i \emph{\textbf{k}} \cdot \bm \delta_1 } + e^{-i \emph{\textbf{k}} \cdot \bm \delta_2 } + \frac{t^\prime}{t} e^{-i \emph{\textbf{k}} \cdot \bm \delta_3 }= 2 \cos(\sqrt{3}k_y a/2) e^{-ik_x a/2} + \frac{t^\prime}{t} e^{i k_x a},
  \end{equation}
  with the NN lattice vectors being defined as $\bm \delta_1 = \frac{a}{2}(1, \sqrt{3})$, $\bm \delta_2 = \frac{a}{2}(1, -\sqrt{3})$, and $\bm \delta_3 = a(-1, 0)$. Accordingly, the dispersion for the normal state is $\epsilon =\pm |\gamma_{\emph{\textbf{k}}}|$.

  \par
  For the critical case of $t^\prime = 2t$, $\epsilon$ vanishes at $\bm M = (\frac{2 \pi}{3  a}, 0)$, and the normal state reaches the semi-Dirac phase. Now we linearize the Hamiltonian $ H$ for a small $\bm k$ around the $\bm M$ point by setting $k_x \rightarrow k_x + \frac{2 \pi}{3 a}$ and $k_y \rightarrow k_y $, with $k_x a \ll 1$ and $k_y a \ll 1$ being satisfied. Up to the second order in $\bm k$, we arrive at
  \begin{subequations}
  \begin{equation}
  \gamma_{\bm M + \bm k} \simeq (3 i a k_x e^{2i \pi /3} + \frac{3}{4} a^2 k^2_y e^{2i \pi /3}),
  \end{equation}
  \begin{equation}
  \gamma^\ast_{\bm M + \bm k} \simeq (-3 i a k_x e^{-2i \pi /3} + \frac{3}{4} a^2 k^2_y e^{-2i \pi /3}),
  \end{equation}
  \begin{equation}
  \gamma_{- \bm M - \bm k} \simeq (-3 i a k_x e^{-2i \pi /3} + \frac{3}{4} a^2 k^2_y e^{-2i \pi /3}),
  \end{equation}
  \begin{equation}
  \gamma^\ast_{-\bm M - \bm k} \simeq (3 i a k_x e^{2i \pi /3} + \frac{3}{4} a^2 k^2_y e^{2i \pi /3}).
  \end{equation}
  \label{eq:gamma}
  \end{subequations}

  \par
  Substituting Eq.~(\ref{eq:gamma}) into Eq.~(\ref{eq:Hk}), we obtain the Hamiltonian at $\bm {q= M + k }$ point as
  \begin{eqnarray}
  \tilde H_{\bm q} &=&  \frac{t}{2} \sum_{\emph{\textbf{q}}, \alpha} [(3ia k_x + \frac{3a^2}{4} k^2_y)e^{\frac{i 2 \pi}{3}} b^\dag_{\emph{\textbf{q}}, \alpha} a_{\emph{\textbf{q}}, \alpha} + (- 3 i a k_x + \frac{3a^2}{4} k^2_y)e^{-\frac{i2 \pi}{3}} a^\dag_{\emph{\textbf{q}}, \alpha} b_{\emph{\textbf{q}}, \alpha} + (- 3ia k_x + \frac{3a^2}{4} k^2_y)e^{-\frac{2 i \pi}{3}} b^\dag_{-\emph{\textbf{q}}, \alpha} a_{-\emph{\textbf{q}}, \alpha} \nonumber\\&& + (3 ia k_x + \frac{3a^2}{4} k^2_y)e^{\frac{2i \pi}{3}} a^\dag_{-\emph{\textbf{q}}, \alpha} b_{-\emph{\textbf{q}}, \alpha}]
  - \frac{\mu}{2} \sum_{\emph{\textbf{q}}, \alpha} [a^\dag_{\emph{\textbf{q}}, \alpha} a_{\emph{\textbf{q}}, \alpha} + b^\dag_{\emph{\textbf{q}}, \alpha} b_{\emph{\textbf{q}}, \alpha} + a^\dag_{-\emph{\textbf{q}}, \alpha} a_{-\emph{\textbf{q}}, \alpha} + b^\dag_{-\emph{\textbf{q}}, \alpha} b_{-\emph{\textbf{q}}, \alpha}] \nonumber\\&&
  + \sum_{\emph{\textbf{q}}, \alpha} [s \Delta (a^\dag_{\emph{\textbf{q}}, \alpha} a^\dag_{-\emph{\textbf{q}}, -\alpha} + b^\dag_{\emph{\textbf{q}}, \alpha} b^\dag_{-\emph{\textbf{q}}, -\alpha})
                                     +s \Delta^\ast (a_{-\emph{\textbf{q}}, -\alpha} a_{\emph{\textbf{q}}, \alpha} + b_{-\emph{\textbf{q}}, -\alpha} b_{\emph{\textbf{q}}, \alpha})].
  \label{eq:Hkl}
  \end{eqnarray}
  where the relation of $\sum_{\bm k} = \sum_{- \bm k}$ has been used. Rewrite $\tilde H_{\bm q}$ in the form of $\tilde H_{\bm q}=\sum_{\bm q} \psi^\dag_{\bm q} H_{\bm q} \psi_{\bm q}$, with the basis
  \begin{equation}
  \psi^\dag_{\bm q} = [a^\dag_{\bm q \uparrow}, b^\dag_{\bm q \uparrow}, a_{-\bm q \downarrow}, b_{-\bm q \downarrow}, a^\dag_{\bm q \downarrow}, b^\dag_{\bm q \downarrow}, a_{-\bm q \uparrow}, b_{-\bm q \uparrow}],
  \end{equation}
  and
  \begin{equation}
  H_{\bm q}= \left( {\begin{array}{*{20}c}
  {\cal H}({\bm k}, \Delta)  & 0  \\
  0 & {\cal H}({\bm k}, -\Delta) \\
  \end{array}} \right).
  \end{equation}
  By defining $v \equiv \frac{3at}{2\hbar}$, $\eta \equiv \frac{3a^2t}{8}$, and $\mu/2 \rightarrow \mu$, the upper block can be formulated as
  \begin{equation}
  {\cal H}({\bm k}, \Delta) = \left( {\begin{array}{*{20}c}
  -\mu & (-i\hbar vk_x+ \eta k^2_y)e^{\frac{-2i\pi}{3}} & \Delta & 0  \\
  (i\hbar vk_x+\eta k^2_y)e^{\frac{2i\pi}{3}} & -\mu & 0 & \Delta \\
  \Delta^\dag & 0 & \mu & (i\hbar vk_x-\eta k^2_y)e^{\frac{-2i\pi}{3}} \\
  0 & \Delta^\dag & (-i\hbar vk_x-\eta k^2_y)e^{\frac{2i\pi}{3}} & \mu \\
  \end{array}} \right),
  \end{equation}
  where the anticommutation relation has been employed. Taking a unitary transformation $H_{\mathrm{BdG}}\equiv {\cal U}^\dag {\cal H} \cal U $ with
  \begin{equation}
  \cal U=
  \left(
    \begin{array}{cccc}
      0 &  e^{\frac{-i\pi}{3}} & 0 & 0 \\
      i e^{\frac{i\pi}{3}} & 0 & 0 & 0 \\
      0 & 0 & 0 & e^{\frac{-i\pi}{3}} \\
      0 & 0 & i e^{\frac{i\pi}{3}} & 0 \\
    \end{array}
  \right).
  \end{equation}
  The BdG Hamiltonian can be compactly written as
  \begin{equation}
  H_{\mathrm{BdG}}= (\hbar v k_x \sigma_x + \eta k^2_y \sigma_y - \mu \sigma_0) \tau_z + \Delta_0 (\cos \phi \sigma_0 \tau_x - \sin \phi \sigma_0 \tau_y),
  \end{equation}
  where we rewrite $\Delta \equiv \Delta_0 e^{i \phi}$ with $\Delta_0$ the amplitude of pairing potential and $\phi$ the superconducting phase, $\sigma_0$ is a $2 \times 2$ identity matrix, and the Pauli matrices $\sigma_{x, y}$ and $\tau_{x, y, z}$ act on the pseudo-spin and Nambu spaces, respectively.

  \end{widetext}
\section{\label{sec:levelappenB}Calculation of the basis scattering states in NS junctions based on semi-Dirac materials}

  In this appendix we present necessary calculation details regarding the wave functions and related quantities in SDMs-based NS junctions.

\subsection{\label{sec:levelappenB1} NS junction extending along the $\textbf{\emph{x}}$-axis}

  For the NS junction extending along the $x$-axis, we assume the translational symmetry is preserved in the $y$ direction, thus the transverse momentum $k_y$ can be treated as a good quantum number. In the S region, solving the BdG equation $H_{\mathrm{BdG}} (-i\partial_x, k_y) \psi = E \psi$ straightforwardly yields
  \begin{subequations}
  \begin{equation}
  \psi_{eq}^\pm =
  \left( {\begin{array} {*{20}c}
  \Lambda^\pm_{eq} \\
  \Gamma_{eq} \\
  \Lambda^\pm_{eq} e^{-i(\beta + \phi)} \\
  \Gamma_{eq} e^{-i(\beta + \phi)}\\
  \end{array}} \right) e^{ \pm i k_{eq} x + i k_y y },
  \end{equation}
  \begin{equation}
  \psi_{hq}^\pm =
  \left( {\begin{array} {*{20}c}
  \Lambda^\pm_{hq} \\
  \Gamma_{hq} \\
  \Lambda^\pm_{hq} e^{i(\beta - \phi)} \\
  \Gamma_{hq} e^{i(\beta - \phi)}\\
  \end{array}} \right) e^{ \pm i k_{hq} x + i k_y y},
  \end{equation}
  \label{eq:wfsx}
  \end{subequations}
  where the related parameters are defined by
  \begin{subequations}
  \begin{equation}
  \hbar v k_{eq(hq)}=+(-)\sqrt{[\mu_S + (-) \Omega]^2- \eta^2 k^4_y },
  \end{equation}
  \begin{equation}
  \Lambda^\pm_{eq(hq)}= \pm \hbar v k_{eq(hq)} - i\eta k^2_y,
  \end{equation}
  \begin{equation}
  \Gamma_{eq(hq)}=\sqrt{\hbar^2 v^2 k^2_{eq(hq)} + \eta^2 k^4_y },
  \end{equation}
  \begin{equation}
  \Omega=\sqrt{E^2-\Delta^2_0},
  \end{equation}
  \end{subequations}

  \par
  In the N region, the basis scattering states can be formulated as
  \begin{subequations}
  \begin{equation}
  \psi^\pm_e= \left( {\begin{array} {*{20}c}
  \pm \hbar v k_e -i \eta k^2_y \\ \varepsilon_+ \\ 0\\ 0\\
  \end{array}} \right) e^{ \pm i k_e x + i k_y y},
  \end{equation}
  \begin{equation}
  \psi^\pm_h= \left( {\begin{array} {*{20}c}
   0\\ 0\\ \mp \hbar v k_h + i \eta k^2_y \\ \varepsilon_- \\
  \end{array}} \right) e^{ \pm i k_h x + i k_y y}.
  \end{equation}
  \label{eq:wfnx}
  \end{subequations}

  \par
  In the interfacial barrier region, the $x$-components of momenta $  k^B_{e(h)} ={\mathrm{sgn}[\varepsilon_{+(-)}]}\sqrt{\varepsilon^2_{+(-)}-\eta^2 k^4_y}/(\hbar v)$. By substituting $\mu_N$ and $k_{e(h)} $, respectviely, with $U$ and $k^B_{e(h)}$ into Eq.~(\ref{eq:wfnx}), we obtain the basis scattering states $\psi^{B\pm}_{e, h}$ in the interfacial barrier region. The related wave functions can be expressed as
  \begin{equation}
  \Psi_{B}= \sum_{m=e, h}{\sum_{n=\pm}{b^n_m \psi^{Bn}_m}},
  \end{equation}
  where $b^n_m$ label the scattering amplitudes.

\subsection{\label{sec:levelappenB2}NS junction along the $y$-axis}

  In the NS junction along the $y$-direction, we assume the transverse momentum $k_x$ is preserved. In the N region, solving the BdG equation $H_{\mathrm{BdG}}(k_x, -i\partial_y) \varphi = E \varphi$ gives the basis states
  \begin{subequations}
  \begin{equation}
  \varphi^\pm_{e,1}= \left( {\begin{array} {*{20}c}
  \hbar v k_x -i \eta q^2_e \\ \varepsilon_+ \\ 0\\ 0\\
  \end{array}} \right) e^{ \pm i q_e y + i k_x x},
  \end{equation}
  \begin{equation}
  \varphi^\pm_{e,2}= \left( {\begin{array} {*{20}c}
  \hbar v k_x +i \eta \kappa^2_e  \\ \varepsilon_+ \\ 0\\ 0\\
  \end{array}} \right) e^{ \mp \kappa_e y + i k_x x},
  \end{equation}
  \begin{equation}
  \varphi^\pm_{h,1}= \left( {\begin{array} {*{20}c}
   0\\ 0\\ -\hbar v k_x + i \eta q^2_h \\ \varepsilon_- \\
  \end{array}} \right) e^{ \pm i q_h y + i k_x x},
  \end{equation}
  \begin{equation}
  \varphi^\pm_{h,2}= \left( {\begin{array} {*{20}c}
   0\\ 0\\ -\hbar v k_x - i \eta \kappa^2_h  \\ \varepsilon_- \\
  \end{array}} \right) e^{ \mp \kappa_h y + i k_x x}.
  \end{equation}
  \label{eq:wfny}
  \end{subequations}

  \par
  In the S region, the basis states can be formulated as
  \begin{subequations}
  \begin{equation}
  \varphi_{eq, 1(2)}^\pm =
  \left( {\begin{array} {*{20}c}
  \lambda_{eq,1(2)} \\
  \gamma_{eq,1(2)} \\
  \lambda_{eq,1(2)} e^{-i(\beta + \phi)} \\
  \gamma_{eq,1(2)} e^{-i(\beta + \phi)}\\
  \end{array}} \right) e^{ \pm i q_{eq, 1(2)} y + i k_x x },
  \end{equation}
  \begin{equation}
  \varphi^\pm_{hq,1(2)} =
  \left( {\begin{array} {*{20}c}
  \lambda_{hq,1(2)} \\
  \gamma_{hq,1(2)} \\
  \lambda_{hq,1(2)} e^{i(\beta - \phi)} \\
  \gamma_{hq,1(2)} e^{i(\beta - \phi)}\\
  \end{array}} \right) e^{ \pm i q_{hq, 1(2)} y + i k_x x},
  \end{equation}
  \label{eq:wfsy}
  \end{subequations}
  with the associated parameters being defined as
  \begin{subequations}
  \begin{equation}
  q_{eq(hq), 1}=+(-)\sqrt{\frac{\sqrt{[\mu_S + (-) \Omega]^2- \hbar^2 v^2 k^2_x}}{\eta}},
  \end{equation}
  \begin{equation}
  q_{eq(hq), 2}=i\sqrt{\frac{\sqrt{[\mu_S + (-) \Omega]^2- \hbar^2 v^2 k^2_x}}{\eta}},
  \end{equation}
  \begin{equation}
  \lambda_{eq(hq), 1}= \hbar v k_x - i \eta q^2_{eq(hq), 1},
  \end{equation}
  \begin{equation}
  \lambda_{eq(hq), 2}= \hbar v k_x - i \eta q^2_{eq(hq), 2},
  \end{equation}
  \begin{equation}
  \gamma_{eq(hq), 1}=\sqrt{\hbar^2 v^2 k^2_x + \eta^2 q^4_{eq(hq), 1} },
  \end{equation}
  \begin{equation}
  \gamma_{eq(hq), 2}=\sqrt{\hbar^2 v^2 k^2_x + \eta^2 q^4_{eq(hq), 2}}.
  \end{equation}
  \end{subequations}
  We note that for $E>\Delta_0$, $ \varphi^\pm_{eq(hq), 1}$ denote the electron(hole)-like scattering states propagating along the $\pm y$-directions, while $ \varphi^\pm_{eq(hq), 2}$ represent the evanescent ones decaying exponentially as $y \rightarrow \pm \infty$. In the case of $E<\Delta_0$, all scattering states given by Eq.~(\ref{eq:wfsy}) describe the evanescent modes.

\vbox{}

%

\end{document}